\documentclass[twocolumn,showpacs,preprintnumbers,amsmath,amssymb]{revtex4-1}

\usepackage{graphicx}
\usepackage{dcolumn}
\usepackage{bm}
\usepackage{color}

\begin{document}

\title{Quantum teleportation over 100 km of fiber using highly-efficient superconducting nanowire single photon detectors}

\author{Hiroki Takesue$^1$} \email{takesue.hiroki@lab.ntt.co.jp}
\author{Shellee D. Dyer$^2$}
\author{Martin J. Stevens$^2$}
\author{Varun Verma$^2$}
\author{Richard P. Mirin$^2$}
\author{Sae Woo Nam$^2$}

\affiliation{
$^1$NTT Basic Research Laboratories, NTT Corporation, 3-1 Morinosato Wakamiya, Atsugi, Kanagawa, 243-0198, Japan\\
$^2$National Institute of Standards and Technology, 325 Broadway, Boulder, CO 80305, USA 
}

\date{\today}



\begin{abstract}
Quantum teleportation is an essential quantum operation by which we can transfer an unknown quantum state to a remote location with the help of quantum entanglement and classical communication. Since the first experimental demonstrations using photonic qubits and continuous variables, the distance of photonic quantum teleportation over free space channels has continued to increase and has reached $>$100 km. On the other hand, quantum teleportation over optical fiber has been challenging, mainly because the multi-fold photon detection that inevitably accompanies quantum teleportation experiments has been very inefficient due to the relatively low detection efficiencies of typical telecom-band single photon detectors. Here, we report efficient quantum teleportation over optical fiber using four high-detection efficiency superconducting nanowire superconducting single-photon detectors (SNSPD) based on MoSi. These SNSPDs make it possible to perform highly-efficient multi-fold photon measurements, allowing us to confirm that the quantum states of input photons were successfully teleported over 100 km of fiber.
\end{abstract}


\maketitle

\section{Introduction}

In the last two decades, we have seen remarkable progress in quantum key distribution (QKD) over optical fiber \cite{gisin,lo}. QKDs over $> 100$ km of fiber have been achieved with schemes based on attenuated laser light \cite{toshiba,takesue,korzh} and entanglement distribution \cite{entqkd}. 
However, to overcome the exponential decrease in the key rate caused by fiber loss and to achieve scalable QKD, we will need a quantum repeater \cite{briegel}, in which quantum teleportation \cite{teleportation} plays a crucial role. 

Since the first experimental demonstrations using photonic qubits \cite{bou} and continuous variables \cite{furusawa} over very short distances on optical tables, the distance of quantum teleportation over free space channels has continued to increase and has reached $>$100 km \cite{ustc,wien}. On the other hand, quantum teleportation over optical fiber has been challenging \cite{marc,buss}, mainly because the multi-fold photon detection that inevitably accompanies quantum teleportation experiments has been very inefficient due to the relatively low detection efficiencies of typical telecom-band single photon detectors. 
Consequently, there have been relatively few reports of quantum teleportation over optical fiber \cite{marc,buss}, and the record distance for quantum teleportation over fiber is 25 km \cite{buss}. Recently, superconducting nanowire single photon detectors (SNSPD) with $>$90\% detection efficiency in the 1.5 $\mu$m band have been realized using superconducting nanowires made of amorphous tungsten silicide \cite{mar}. 
In the present work, we employed SNSPDs based on another amorphous material, molybdenum silicide (MoSi) to perform photonic quantum teleportation over fiber. We used four high detection efficiency (80-86\%) MoSi SNSPDs, which enabled us to perform highly-efficient multi-fold coincidence measurements, which led to successful quantum teleportation over 100 km of fiber.

\begin{figure*}[t]
\centerline{\includegraphics[width=0.8\linewidth]{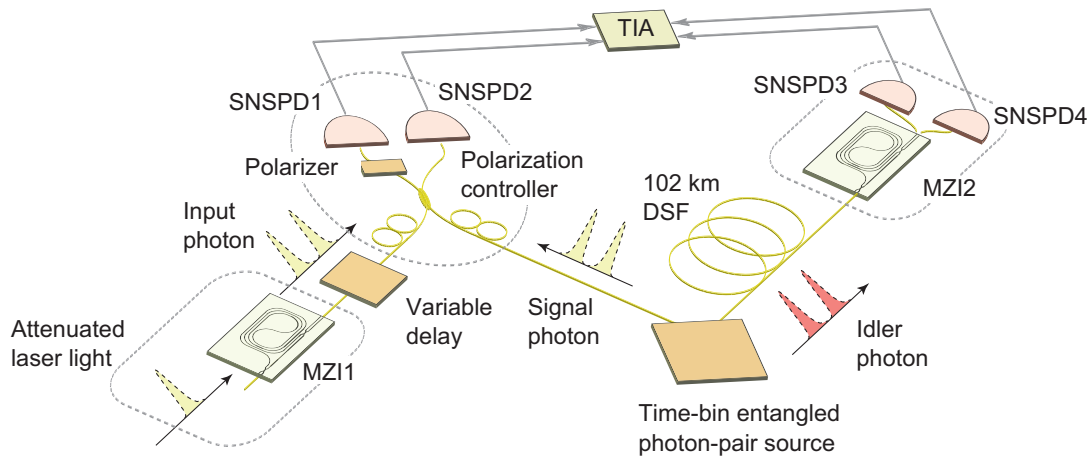}}
\caption{Experimental setup. Yellow and grey solid lines indicate the optical fibers and electrical lines, respectively. SNSPD: superconducting nanowire single photon detector, MZI: delayed Mach-Zehnder interferometer, DSF: dispersion shifted fiber, TIA: time interval analyzer. }
\label{setup}
\end{figure*}

\section{Implementation of quantum teleportation}

\subsection{Quantum teleportation with time-bin qubits}
As a quantum information carrier, we use a photon encoded as a time-bin qubit, which is a coherent superposition of two temporal modes \cite{timebin}. A photonic time-bin qubit is more suitable for fiber transmission than a polarization qubit, because it is generally difficult to preserve a polarization state in a long fiber. 
We assume that the quantum state of the input photon, namely the photon whose quantum state is to be teleported, is given by $\alpha |1\rangle_{in} +\beta |2\rangle_{in}$, where $\alpha$ and $\beta$ are complex numbers that satisfy $|\alpha|^2 + |\beta|^2 = 1$, and $|x\rangle_y$ denotes a state where there is a photon at a time slot $x$ in a mode $y$. 
The quantum state of the time-bin entangled photon pairs that we use in our experiment is expressed as
\begin{equation}
|\psi\rangle = \frac{1}{\sqrt{2}}(|1\rangle_s |1\rangle_i + |2\rangle_s |2\rangle_i), \label{1}
\end{equation}
where subscripts $s$ and $i$ denote the signal and idler modes, respectively. 
Using Eq. (\ref{1}), the state of the whole system is expressed as
\begin{eqnarray}
& & \frac{1}{\sqrt{2}}(|1\rangle_s |1\rangle_i + |2\rangle_s |2\rangle_i)(\alpha |1\rangle_{in} + \beta |2\rangle_{in}) \nonumber \\
&=& \frac{1}{2} \left[ 
|\Phi^+\rangle (\alpha |1\rangle_i + \beta |2\rangle_i) + |\Phi^-\rangle (\alpha |1\rangle_i -\beta |2\rangle_i) \right. \nonumber \\
& & \left. + |\Psi^+\rangle (\beta |1\rangle_i + \alpha |2\rangle_i) + |\Psi^-\rangle (\beta |1\rangle_i - \alpha |2\rangle_i)
\right], \label{tel}
\end{eqnarray}
where $|\Psi^\pm\rangle$ and $\Phi^\pm\rangle$ are the Bell states for the signal and input photons given by
\begin{eqnarray*}
|\Psi^\pm\rangle &=& \frac{1}{\sqrt{2}} (|1\rangle_s |2\rangle_{in} \pm |2\rangle_s |1\rangle_{in}), \\
|\Phi^\pm\rangle &=& \frac{1}{\sqrt{2}} (|1\rangle_s |1\rangle_{in} \pm |2\rangle_s |2\rangle_{in}).
\end{eqnarray*}
The right hand side of Eq. (\ref{tel}) shows that the amplitudes of the input photon $\alpha$ and $\beta$ are transferred to that of the idler photon with some unitary transformation depending on the results of Bell state measurements. 
We implement Bell state measurements using a beamsplitter followed by single photon detectors without a photon number resolving capability as in \cite{marc}.  With this scheme, $|\Phi^\pm\rangle$ cannot be observed because the two photons are always output from the same port of the beamsplitter due to the HOM effect \cite{hom}. When the state $|\Psi^+\rangle$ is input into a beamsplitter, the two photons are output from the same port but in different time slots, so in theory this event can be observed by detecting consecutive clicks in the same detector. However, in our experiment, the SNSPD recovery time is $\sim 100$ ns and consecutive photon pulses cannot be detected, and thus projection to the $|\Psi^+\rangle$ state cannot be implemented. On the other hand, the $|\Psi^-\rangle$ state launched into a beam splitter gives a coincidence between the two detectors at different time slots, which means that only this state can be distinguished. As a result, we can teleport the quantum state of the input photon when the signal and input photons are projected onto the $|\Psi^-\rangle$ state with a probability of 1/4.

According to the last term on the RHS of Eq. (\ref{tel}), the states of input photon $|1\rangle_{in}$, $|2\rangle_{in}$, $|L\rangle_{in}$, $|R\rangle_{in}$, and $|\pm\rangle_{in}$ are transformed to $|2\rangle_i$, $|1\rangle_i$, $|L\rangle_i$, $|R\rangle_i$, and $|\mp\rangle_i$, respectively.

\subsection{Time-bin entangled photon pair source}

Here we describe our time-bin entanglement source. Pulses from a fiber mode-locked laser are passed through a wavelength filter whose center wavelength and bandwidth are 1551.1 nm and 20 GHz, respectively. As a result, we obtain pulses with a temporal width of $\sim$20 ps. The pulses are launched into a delayed Mach-Zehnder interferometer whose delay time is 1 ns to create double pulses. The double pulses are amplified by an erbium-doped fiber amplifier (EDFA) and launched into the first periodically poled lithium niobate (PPLN) waveguide, where a pump pulse train with a wavelength of 775.6 nm is generated via the second harmonic generation process. The 775.6-nm pulses are then passed through a filter to eliminate the 1551.1-nm light and input into the second PPLN waveguide, where time-bin entangled photon pairs whose state is given by Eq. (\ref{1}) are generated through the SPDC process. 
A portion of the output from the fiber mode-locked laser is extracted by a 99:1 fiber coupler, passed through an attenuator to make the average photon number per pulse much less than 1, and then launched into an optical bandpass filter whose wavelength and transmission bandwidth are exactly the same as those used for separating signal photons. The output attenuated laser pulse is used for preparing input qubits. 
The configuration of the time-bin entangled photon pair and the attenuated laser pulse preparation are described in detail in Supplement 1. 

\subsection{Quantum teleportation setup}
The setup for our quantum teleportation experiment is shown in Fig. \ref{setup}.  
A time-bin entanglement source described above generates entangled photon pairs in the 1.5-$\mu$m band at a clock frequency of 35.53 MHz. 
The generated photon pairs are separated into signal (1546.3 nm) and idler (1555.9 nm) channels using a wavelength filter with a transmission bandwidth of 20 GHz for both signal and idler channels, resulting in a photon pair coherence time of $\sim$20 ps. 
An attenuated laser pulse described in the previous subsection is input into a delayed Mach-Zehnder interferometer (MZI1) whose delay time is 1 ns to create a time-bin qubit for use as an input photon.  Our delayed interferometers are based on a silica waveguide. The phase differences between the two waveguide arms are stably controlled by tuning the waveguide temperature \cite{honjo}.


The signal and input photons are passed through polarization controllers and then launched into a 3-dB fiber coupler followed by SNSPD1 and 2. Here, the temporal distinguishability between the two photons is eliminated by adjusting the temporal position of the input photon using a variable delay line, while the polarization distinguishability is eliminated by adjusting the polarization controllers to maximize the number of photons that passed through a polarizer placed in front of SNSPD1. We can project the two photons into $|\Psi^-\rangle$ by conditioning on events where both SNSPD1 and 2 detect photons, but in different time slots, as described above. The idler photon is transmitted over 102 km of dispersion shifted fiber, and is received by the second delayed interferometer (MZI2) whose two output ports are connected to SNSPD3 and 4. Thanks to the narrow bandwidth of the idler photon, the temporal waveform broadening caused by higher-order dispersion in a fiber is negligibly small. 
The detection signals from the SNSPDs are received by a time interval analyzer (TIA) for coincidence measurements. 
By conditioning the detection events at SNSPD3 or 4 on the successful projection of the signal and input photons to $|\Psi^-\rangle$, we can teleport the quantum state of an input photon to that of the idler photon with a pre-determined unitary transformation. We set the average photon number per qubit for the entangled photon pairs and the input photons at 0.016.

It should be noted that we can implement two non-orthogonal measurements using MZI2 followed by two single photon detectors \cite{tittel,tomo}. Let us denote the phase difference between the two arms of MZI2 by $\theta_2$. When a time-bin qubit passes through MZI2, the detectors can possibly detect a photon in three time slots. The detection of a photon in the second time slot with SNSPD3 and 4 means that the input qubit is projected to the states $(|1\rangle + e^{i \theta_2} |2\rangle)/\sqrt{2}$ and $(|1\rangle - e^{i \theta_2} |2\rangle)/\sqrt{2}$, respectively. On the other hand, when we detect a photon in the first or third time slots with either of the detectors, the qubit is projected to state $|1\rangle$ and $|2\rangle$, respectively. Hereafter, we refer to a projection measurement at the second slot as an energy basis measurement, and that at the first or third slots as a time basis measurement.


\section{MoSi SNSPD}

\begin{figure}[thb]
\centerline{\includegraphics[width=\linewidth]{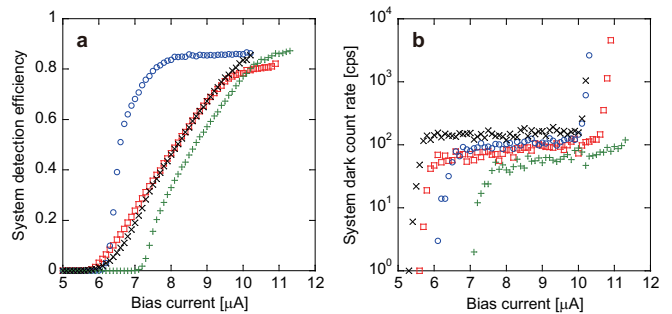}}
\caption{System detection efficiency (a) and dark count rate (b) of the SNSPD system. Squares: SNSPD1, circles: SNSPD2, crosses: SNSPD3, and x symbols: SNSPD4. }
\label{detector}
\end{figure}

The SNSPDs are based on a 15 $\mu$m $\times$ 15 $\mu$m MoSi meander whose nominal wire width and pitch are 160 and 210 nm, respectively. These devices are placed in a cryostat and cooled to 0.7 K. A 3.5-cm diameter, 5-turn fiber coil, which works as a blackbody filter, efficiently eliminates the photons whose wavelengths are longer than 2 $\mu$m \cite{varun}. The coil is formed in a fiber pigtail in front of each SNSPD inside the cryostat. 
The system detection efficiencies and background count rates as a function of bias current are plotted in Fig.\ref{detector} (a) and (b), respectively. We obtained $>$80\% detection efficiencies for all four detectors used in the experiment. In the following experiment, the bias currents of the SNSPDs were set to give system detection efficiencies of 80\% (SNSPD1), 86\% (SNSPD2 and 3) and 81\% (SNSPD4). With these detection efficiencies, the background count rates were $\sim 10^2$ cps for all the detectors, which is an order of magnitude smaller than that reported in \cite{mar} ($\sim 10^3$ cps) thanks to the blackbody filter. In addition, the timing jitter of the SNSPDS was $\sim90$ ps, and thus the measurement errors caused by inter-symbol interference were negligible in our experiment where the temporal difference between time bins was 1 ns. 
We also observed that the polarization dependence of the detection efficiency was less than 5\% for all the detectors. This polarization insensitivity is probably due to the larger fill factor of the meander, which is achieved by designing the wire width and pitch so that their sizes are similar. The polarization insensitivity of the SNSPDs contributed to the stable multi-fold coincidence measurement for a long time in our long-distance quantum teleportation experiment described below. The polarization dependence of the SNSPDs will be reported in detail elsewhere.

\begin{figure*}[thb]
\centerline{\includegraphics[width=.9\linewidth]{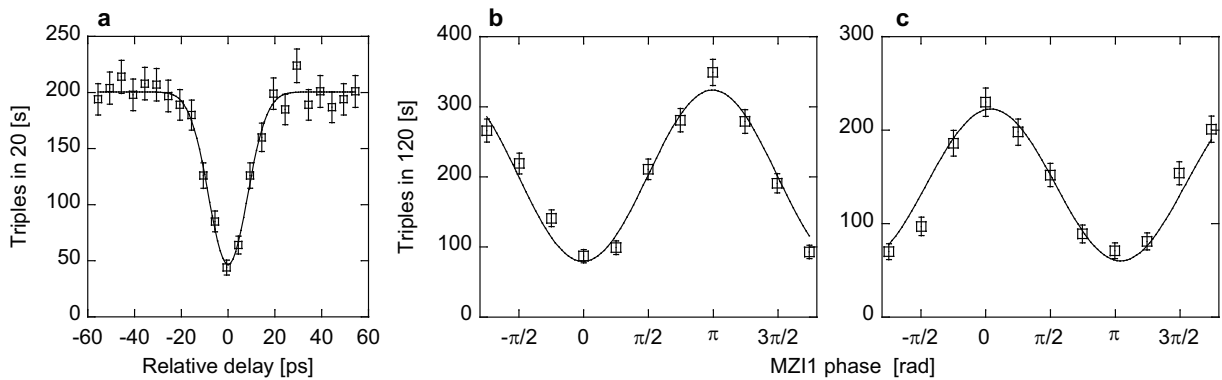}}
\caption{Experimental results for (a) HOM interference and (b), (c) quantum teleportation without transmission fiber. (b) and (c) correspond to the number of triple coincidences in 120 s as a function of MZI1 phase observed by SNSPD3 and 4, respectively. }
\label{hom}
\end{figure*}

\begin{figure*}[htb]
\centerline{\includegraphics[width=.9\linewidth]{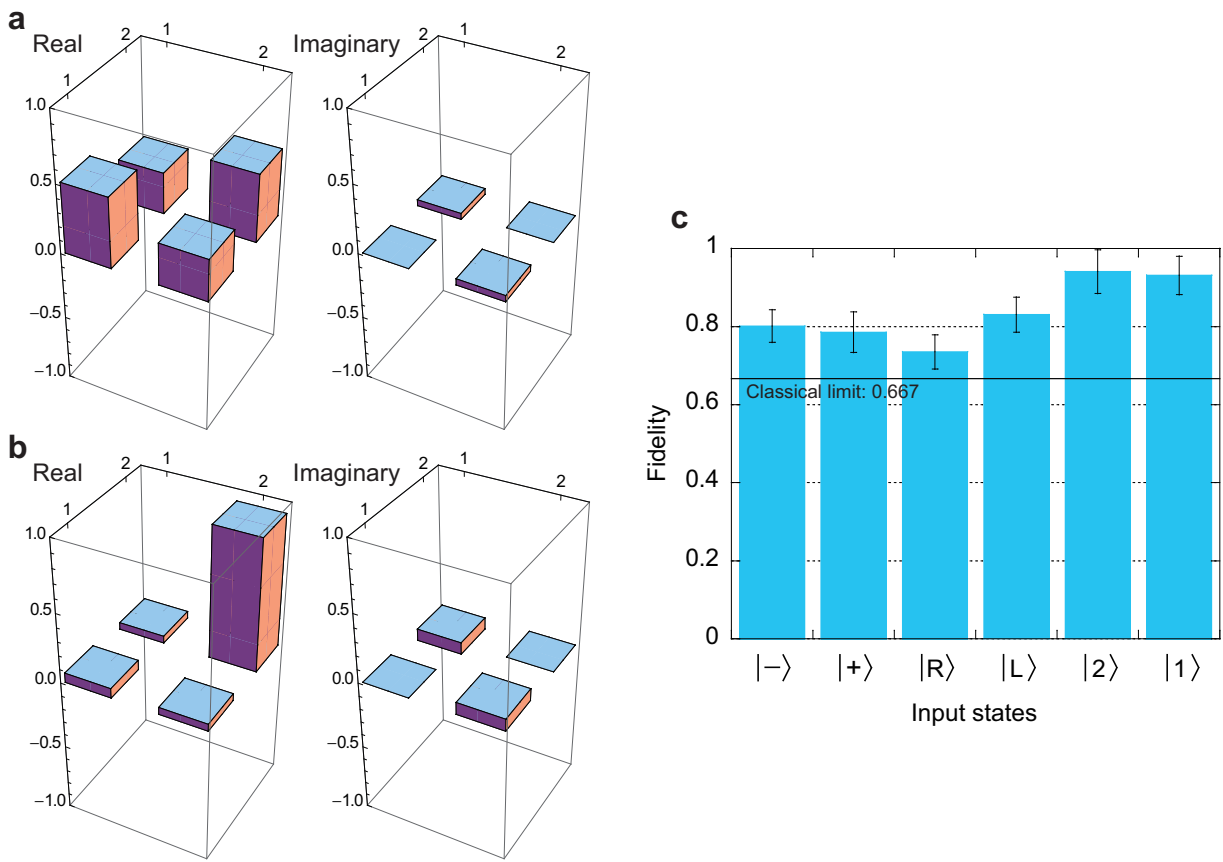}}
\caption{Results of quantum teleportation over 100 km of fiber. (a) Density matrices obtained by QST and maximum likelihood estimation for target states $|+\rangle$ and (b) $|2\rangle$. (c) Experimentally-obtained fidelities for six distinct input states.}
\label{100km}
\end{figure*}

\section{Experimental results}

\subsection{Hong-Ou-Mandel interference}

We performed a Hong-Ou-Mandel (HOM) interference experiment to investigate the indistinguishability of the input and the signal photons \cite{hom}. We removed the transmission fiber and MZIs, and the idler photons were directly received by SNSPD3. 
We observed the HOM interference by taking the triple coincidences between SNSPD1, 2 and 3 as a function of relative delay between the input and signal photons. Note that this experiment corresponds to quantum interference between an attenuated laser light and a heralded single photon \cite{rarity}. 
The result is shown in Fig. \ref{hom}(a). At zero relative delay, we observed a clear dip in the triples. The visibility of the HOM dip was $76.9 \pm 3.4$\%, which significantly exceeded the classical limit of HOM visibility of 50\%.


\subsection{Quantum teleportation}

We then performed a quantum teleportation experiment. We first removed the transmission fiber from the setup shown in Fig. \ref{setup}. We launched input photons whose state is given by $(|1\rangle + e^{i \theta_1}|2\rangle)\sqrt{2}$ by setting the MZI1 phase at $\theta_1$, and observed the photon detection events with SNSPD3 and 4 in the energy basis, conditioned on the projection of the signal and input photons to $|\Psi^-\rangle$. The result is shown in Fig. \ref{hom} (a) (SNSPD3) and (b) (SNSPD4). Thus, we confirmed that the phase of the input photon was successfully transferred to that of the idler photon. The visibilities were $60.5\pm 5.0$\% for the SNSPD3 channel and $57.5 \pm 5.7$\% for the SNSPD4 channel. Note that we did not subtract any accidental coincidences or background counts from any of the experimental data shown in this paper.


We then performed a quantum teleportation experiment over 100 km of fiber for six distinct input states. We prepared states $|\pm\rangle = (|1\rangle \pm |2\rangle)/\sqrt{2}$, $|L\rangle = (|1\rangle +i|2\rangle)/\sqrt{2}$, and $|R\rangle = (|1\rangle -i|2\rangle)/\sqrt{2}$) by adjusting the MZI1 phase $\theta_1$, while the other two states $|1\rangle$ and $|2\rangle$ were obtained by removing MZI1 and adjusting the temporal positions of the qubits so that they coincided with that of the first and second pulses of the signal photons, respectively. We undertook quantum state tomography (QST) \cite{james} on the teleported states to obtain their density matrices. 
We employed six projection measurements that correspond to $|\pm\rangle$, $|L\rangle$, $R\rangle$, $|1\rangle$ and $|2\rangle$ to perform QST on a single qubit. 
The projections to $|+\rangle$ and $|-\rangle$ were achieved by performing on energy basis measurement with $\theta_2=0$ at SNSPD3 and 4, respectively, while $|L\rangle$ and $|R\rangle$ projections were implemented by employing energy basis measurements at SNSPD3 and 4, respectively, with $\theta_2=\pi/2$. The projection to $|1\rangle$ and $|2\rangle$ were achieved with time basis measurements at the first and third time slots, respectively.  
Since our setup was inherently equipped with two non-orthogonal measurement bases, we could perform the six projections with only two TIA measurements at $\theta_2=\{0,\pi/2\}$ \cite{tomo}. 
The data acquisition time for each TIA measurement was 6000 s, which means that the total QST measurement time for each input state was 12000 s. The average number of triple coincidences for each basis was $\sim$170. With the raw data obtained in the QST measurements, we performed a maximum likelihood estimation to obtain physically legitimate matrices \cite{james}. 
The reconstructed density matrices after the maximum likelihood estimation for the target states $|+\rangle$ and $|2\rangle$ are shown in Fig.\ref{100km} (a) and (b) (the matrices for the other four target states are provided in Supplement 1). 
Using these density matrices, we calculated the fidelity for each input state. The results are shown in Fig. \ref{100km} (c). We obtained an average fidelity of $83.7 \pm 2.0$\% for the six distinct input states, which means that we observed a violation of the classical limit (66.7\%) by more than eight standard deviations. 

\section{Summary}
In summary, we have demonstrated the quantum teleportation of a photonic qubit over 100 km of optical fiber. This result confirmed the feasibility of long-distance quantum communication based on quantum teleportation over optical fiber. 
In addition, we can expect that the highly-efficient multi-fold photon measurement using the SNSPDs will pave the way towards advanced quantum communication systems based on multi-photon quantum states such as the Greenberger-Horne-Zeilinger state \cite{pan} and the cluster state \cite{cluster} over optical fiber.



\section*{Acknowledgements}
We thank Thomas Gerrits, Robert Horansky, and Edward Tortorici for fruitful discussions.





\vspace{5mm}
\noindent
See Supplement 1 for supporting content

\end{document}